\input{aipcheck}

\documentclass[
    ,final            
  ]
  {aipproc}

\layoutstyle{6x9}

\begin{document}

\title{Lattice QCD simulations with light dynamical quarks}


\author{Sinya AOKI}{
  address={Graduate School of Pure and Applied Sciences, University of
  Tsukuba, Tsukuba, \\ Ibaraki 305-8571, Japan}
  ,altaddress={Riken BNL Research Center, Physics 510A, BNL, Upton,
  NY11973, USA} 
}

\begin{abstract}
I report recent results from full QCD simulations by CP-PACS and JLQCD collaborations.
\end{abstract}

\maketitle


\section{Introduction}
Lattice QCD is a powerful tool to understand the strong interaction of
hadrons from the first principles of QCD for quarks and gluons with the
aid of numerical simulations. Physical quantities calculated with
the method range from the spectrum of light hadrons to electroweak 
matrix elements. Systematic errors such as finite lattice volume and
spacing, and the use of the quenched approximation are gradually being
reduced thanks to development of computer power as well as simulation
algorithms. Among these systematics, a current main concern is the
effect of light dynamical quarks.

As a member of CP-PACS and JLQCD collaborations in Japan, I have been
working on large scale lattice QCD simulations for many years. In this talk I
report recent results of our collaborations
in lattice QCD simulations with light dynamical quarks.

\begin{figure}
  \includegraphics[height=.25\textheight]{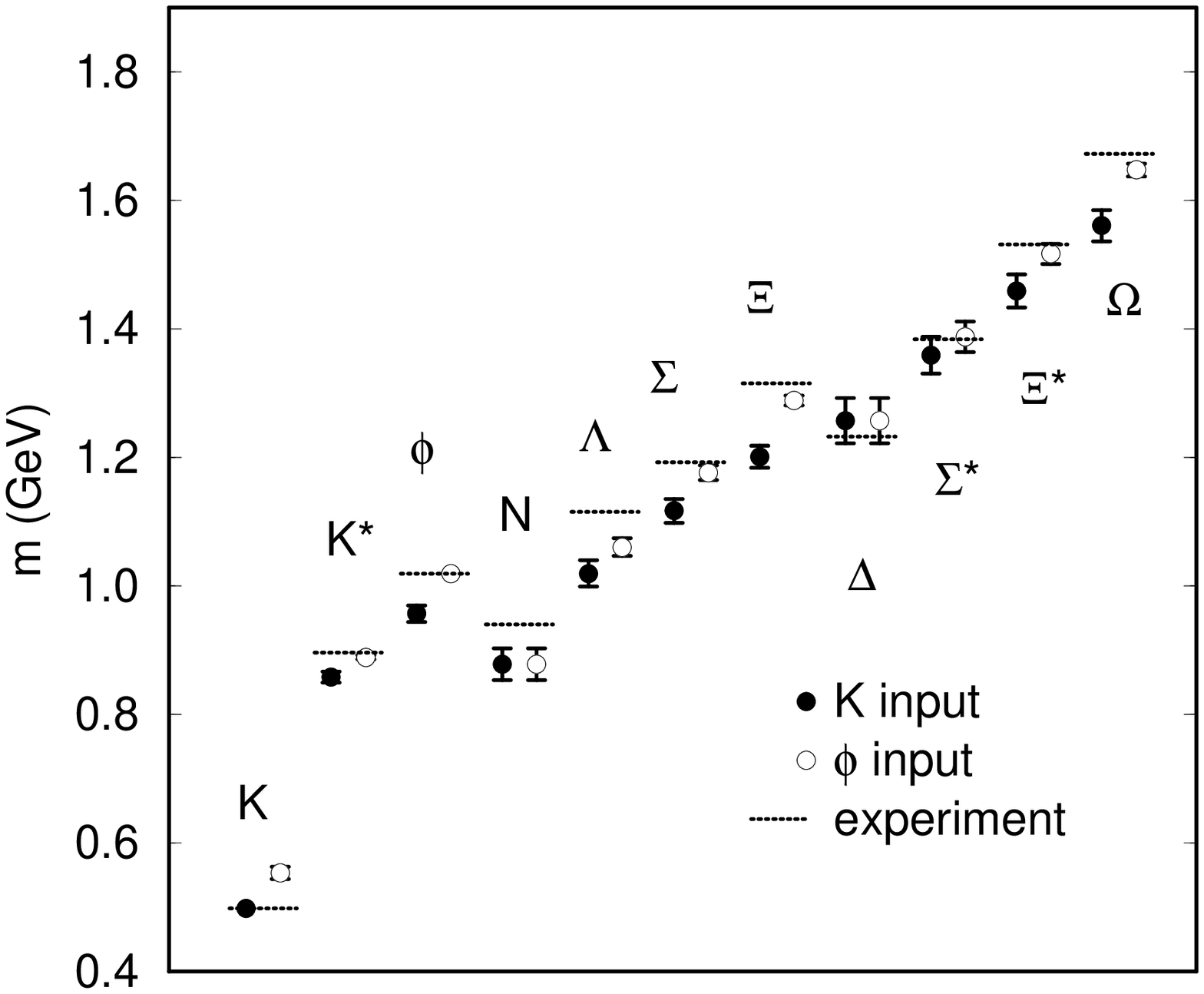}
  \includegraphics[height=.25\textheight]{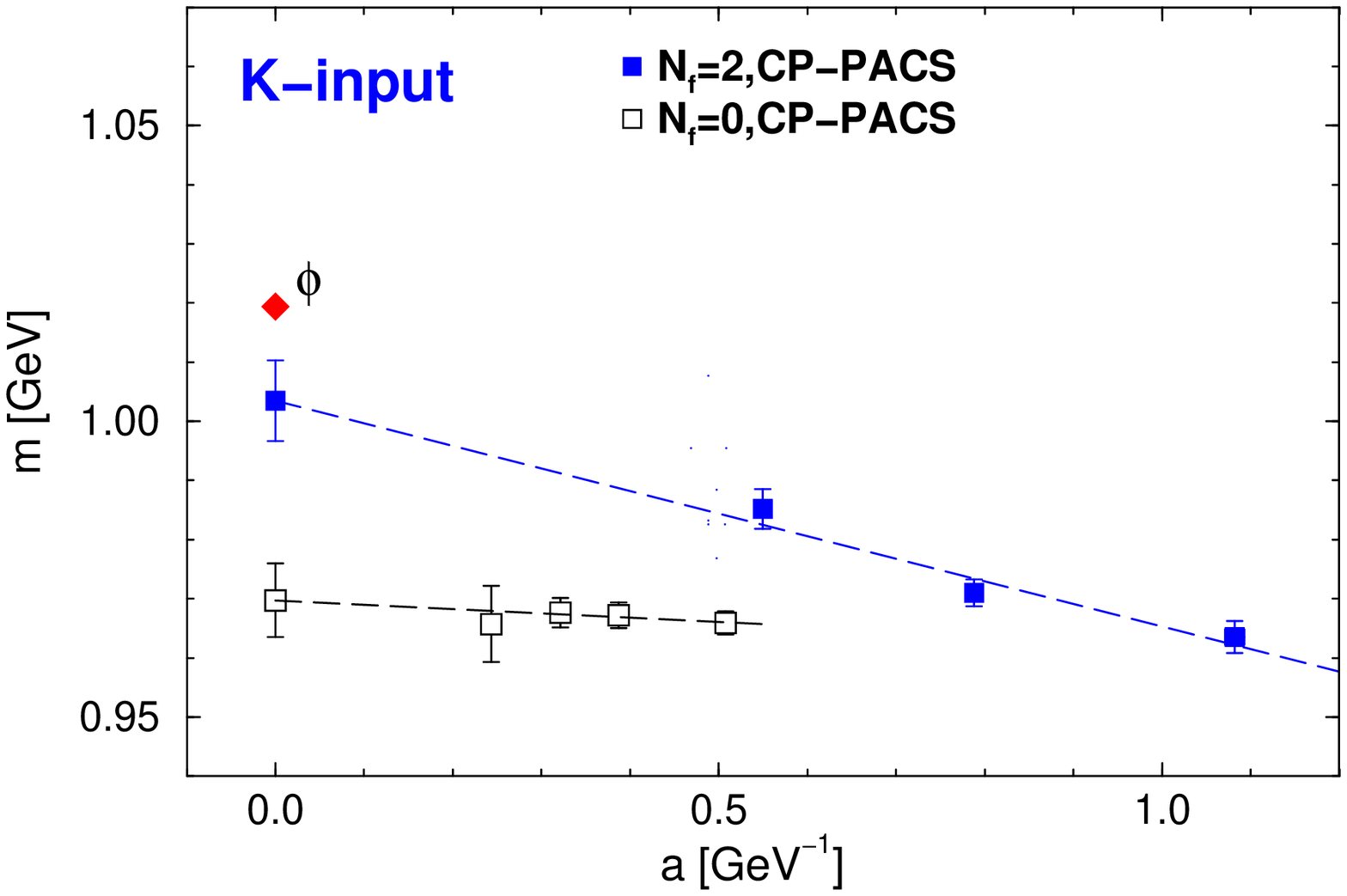}
  \caption{Left: Light hadron spectra in quenched QCD.
  Right: $\phi$ meson masses from $K$ input
as a function of $a$ in $N_f=2$ full QCD (solid), 
together with the quenched results (open).}
  \label{fig:spectra}
\end{figure}

Let me first explain the necessity of dynamical quark effects,  by
presenting the quenched light hadron spectrum  in the left panel of 
Fig.~\ref{fig:spectra}. 
These results have been obtained by the CP-PACS collaboration
after taking the continuum limit\cite{cppacs1}.  In this calculation the experimental $\rho$ and $\pi$ meson masses are used to fix the lattice spacing $a$ and the light quark mass $m_l$, where
up and down quark masses are assumed to be equal ($m_u=m_d = m_l$).
For the strange quark mass, two choices are compared, one employing 
the $K$ meson mass (filled symbols ; $K$-input) and other with 
the $\phi$ meson mass (open symbols; $\phi$-input). 
Experimental values are given by horizontal lines.
This figure shows an overall agreement of the light hadron spectrum 
in the quenched lattice QCD at a 5--10\% level. 
However, it is also clear that there are systematic deviations between 
the quenched spectrum and experiments beyond statistical errors 
of 2--3\%. In particular, the hyperfine splitting
between the $\phi$, $K^*$ meson masses and the $K$ meson mass is smaller than
the experimental one.
This indicates that full QCD simulations are indeed necessary for more accurate results.
We then performed a large scale 2 flavor full QCD 
simulations\cite{cppacs2}.
The right panel of Fig.~\ref{fig:spectra} shows the $\phi$ meson mass
from the $K$ input as a function of the lattice spacing $a$.
As can been seen from the figure, the deviation from an experimental value in quenched QCD 
is much reduced in the $N_f=2$ full QCD after the continuum extrapolation.
The effect of dynamical sea quarks is really
important for reproducing the correct spectrum.

\section{$N_f=2+1$ full QCD simulations}
The success of 2 flavor full QCD simulations motivates us to perform more accurate calculations, $N_f=2+1$ flavor full QCD simulations, in order to remove the systematic error associated with the absence of the dynamical strange quark.

\subsection{$N_f=2+1$ full QCD project}
We have started 2+1 full QCD simulations as a joint project of CP-PACS and JLQCD collaborations
\cite{ishikawa1,ishikawa2,ishikawa3}, 
employing the RG improved gauge action and
the Wilson-type quark action.
In order to reduce the effect of the explicit chiral symmetry violation 
in the Wilson quark, 
we introduce the non-perturbative $O(a)$ improvement. A necessary parameter $c_{\rm SW}$ has already been determined by our collaborations\cite{joint1}, prior to  large scale simulations.
We employ the standard Hybrid Monte-Carlo (HMC) algorithm to simulate degenerate up and down quarks, while polynomial HMC algorithm for the dynamical strange quark. The latter algorithm has 
been developed by us to simulate odd number of dynamical quarks\cite{joint2}. 

\subsection{Simulations and analyses}
We take 3 values of lattice spacing, $a\simeq$ 0.07, 0.10, 0.12 fm, 
equally spaced in $a^2$,  
in order to perform the continuum extrapolation, with the (2 fm)$^3$
spatial volume. 
We accumulate more than 5,000 HMC trajectories at each lattice spacing.
We take 5 values of the degenerate up and down quark mass ranging between 
$m_{\rm PS}/m_{\rm V} \simeq 0.6$ and 0.78, where $m_{\rm PS}$ and $m_{\rm V}$ are
 the pseudo-scalar meson mass and the vector meson mass, respectively. 
 For the strange quark mass, we take 2 values
around   $m_{\rm PS}/m_{\rm V} \simeq 0.7$. Note that our light quark
mass is much heavier than the experimental value, $m_{\rm PS}/m_{\rm V}
=0.18$, while the strange quark mass is close to the value, $m_{\rm
PS}/m_{\rm V} \simeq 0.68$, 
estimated by the 1-loop chiral perturbation theory.

For the chiral extrapolation of meson masses, we have used polynomial functions in quark masses, including up to quadratic terms with an interchange symmetry among 3 sea quarks and that among
2 valence quarks. Chiral fits are made for light-light(LL),
light-strange(LS) and strange-strange(SS) mesons
simultaneously. Polynomial functions describe 
quark mass dependences of data very well\cite{ishikawa3}.

In order to estimate the systematics of the polynomial chiral extrapolation,
we also employ another fit function, obtained by the Wilson
chiral perturbation theory (WChPT)\cite{aoki1,chpt1,chpt2}, which contains both chiral loop and finite lattice spacing effects.
No difference between the WChPT fit and the polynomial fit is observed
for PS meson masses, while 
a slight difference is detected in the small quark mass region for 
V meson masses\cite{ishikawa3}.
This analysis suggests that the effect of chiral log is small in the region of the light quark mass employed in our simulations.
Note however that our light quark mass may be too heavy to apply the NLO formula. 
Further analysis including data with lighter quark mass will be required for the definite conclusion on the effect of the chiral log to meson masses.

\subsection{Continuum extrapolation}
\begin{figure}
  \includegraphics[width=\textwidth, angle=0]{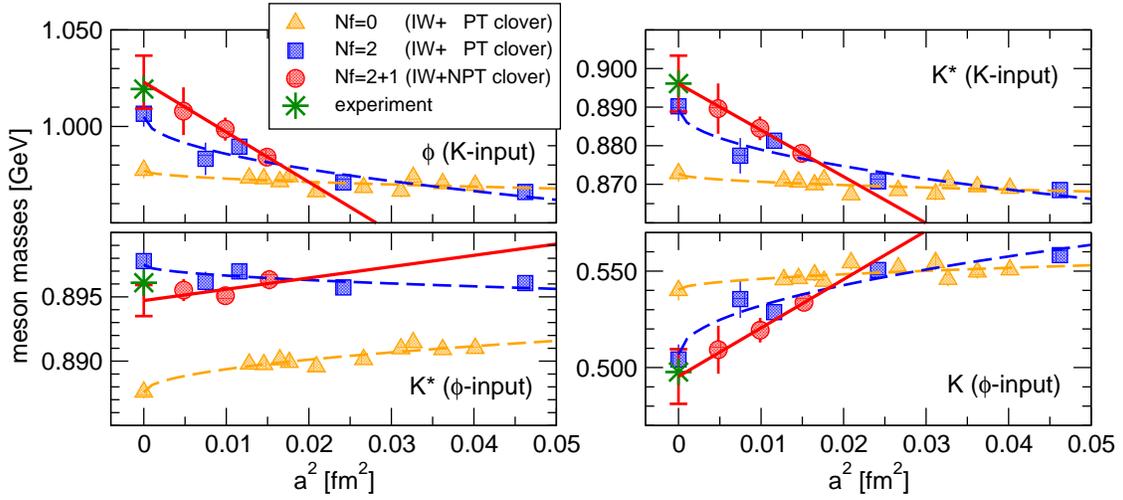}
  \caption{Left-Top: $\phi$ meson mass as a function of $a^2$ with $K$ input.
  Left-Bottom: $K^*$ meson mass with $K$ input. Right-Top: $K$ meson mass with $\phi$ input.
  Right-Bottom: $K^*$ with $\phi$ input.}
  \label{fig:cont_mass}
\end{figure}
Now let me consider the continuum extrapolation of some quantities.

\subsubsection{Light meson spectra}
Left two panels of Fig.\ref{fig:cont_mass} show vector ($\phi$ and
$K^*$) meson masses
as the function of $a^2$ with the strange quark mass from the $K$ input,
while right panels are  $K^*$ and $K$ meson masses with the strange
quark mass from the $\phi$ input. Circles represent 2+1 flavor full QCD
results, while 2 flavor and quenched results are given for comparison by 
squares and triangles, respectively. 
2+1 flavor results are consistent with experimental values within 2\% statistical errors after the continuum extrapolation. This agreement is encouraging.
It is difficult, however, to pin down the effect of the dynamical
strange quark on meson spectra, since 2\% errors are much larger than
those of 2 flavor results. 

\subsubsection{Quark mass}
Quark masses are determined for the $\overline{\mbox{MS}}$ 
scheme at the scale $\mu=2$~GeV. 
Lattice results are translated to the $\overline{\mbox{MS}}$ scheme
at $\mu=a^{-1}$ using tadpole-improved one-loop matching 
factor~\cite{Aoki:1998-PT-renorm}, and then evolved to 
$\mu=2$~GeV using the four-loop RG-equation.
\begin{figure}
  \includegraphics[width=\textwidth, angle=0]{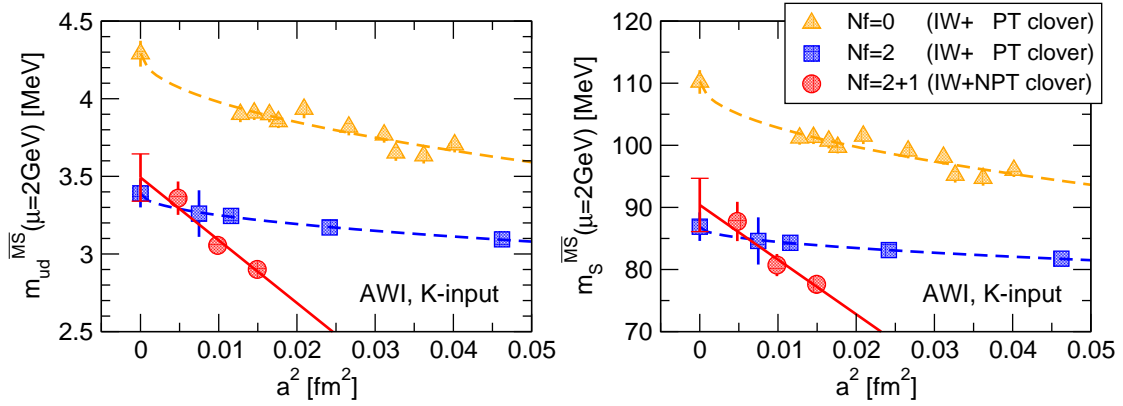}
  \caption{Left: The light quark mass $m_{\rm ud}$  as a function of $a^2$. 
  Right: The strange quark mass $m_{\rm s}$. }
  \label{fig:cont_qmass}
\end{figure}
Quark mass results are shown in Fig.~\ref{fig:cont_qmass}.
As already observed in $N_f=2$ QCD~\cite{cppacs2}, 
values of the strange quark mass determined for either the $K$- or 
the $\phi$-inputs, while different at finite lattice spacings, extrapolate 
to a common value in the continuum limit. 
Therefore the quark masses in the continuum limit is estimated from 
a combined fit to data with the $K$- and the $\phi$-inputs.
We finally obtain\cite{ishikawa3}
\begin{equation}
 m_{\rm ud}^{\overline{\rm MS}}(\mu=2~{\rm GeV})
=3.50(14)({}^{+26}_{-15})~{\rm MeV},\;\;\;\;\;
 m_{\rm s}^{\overline{\rm MS}}(\mu
=2~{\rm GeV})=91.8(3.9)({}^{+6.8}_{-4.1})~{\rm MeV}.
\end{equation}
Dynamical up and down quarks reduce significantly the quark 
masses.
The effect of strange quark is less dramatic, and we do  not see 
deviations from the $N_f=2$ results, $m_{\rm ud} = 3.44^{+0.14}_{-0.22}$ Mev, 
$m_{\rm s} = 88^{+6}_{-6}$ Mev ($K$ input)\cite{cppacs2} 
beyond statistical errors.

\subsubsection{Pseudo-scalar meson decay constant}
PS meson decay constants are estimated using matching factor
determined by tadpole-improved one-loop perturbation theory.
The results with the $K$-input are
\begin{equation}
f_{\pi}=140.7(9.3)~{\rm MeV},\;\;\;
f_K=160.9(9.1)~{\rm MeV},\;\;\;
f_K/f_{\pi}=1.142(17).
\end{equation}
We recall that in our $N_f=2$ QCD calculation, 
the magnitude of scaling violation was so large that we were not able 
to estimate values in the continuum 
limit~\cite{cppacs2}.
The situation is much better in the present case and $f_{\pi}$ and $f_K$ turn
out to be consistent with experiment.  The errors are large, however.  
Furthermore, the ratio $f_K/f_{\pi}$ differs significantly from the 
experimental value, 1.223(12).
A long chiral extrapolation is a possible cause of the discrepancy. 

\subsection{Flavor singlet mesons}
\begin{figure}
  \includegraphics[width=.5\textwidth, angle=0]{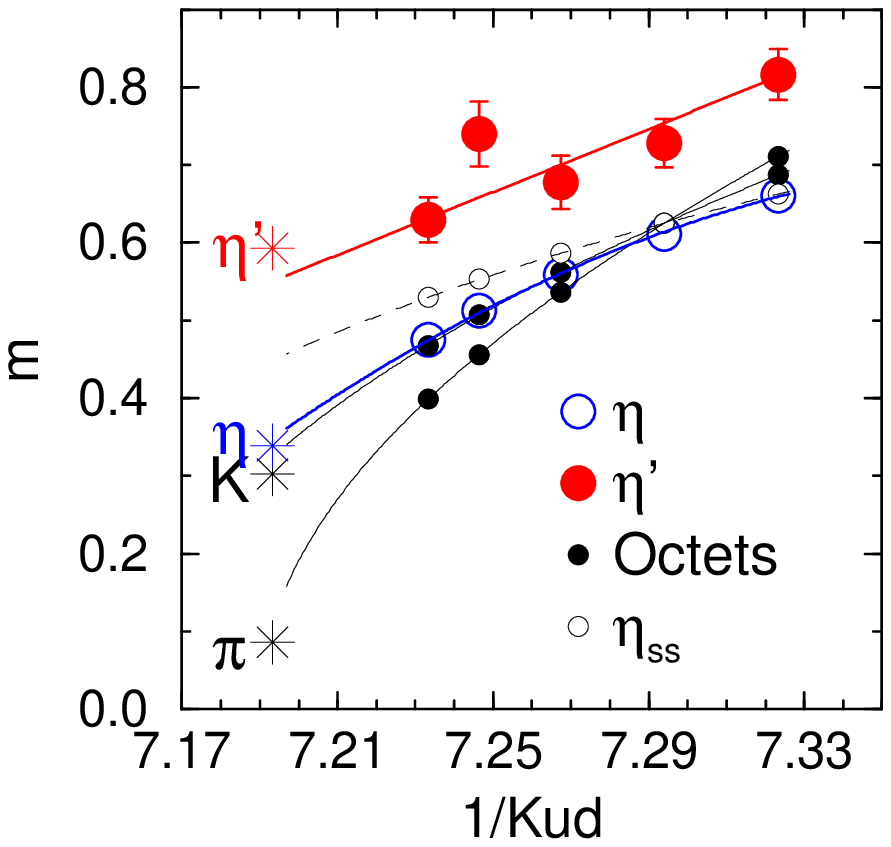}
  \includegraphics[width=.5\textwidth, angle=0]{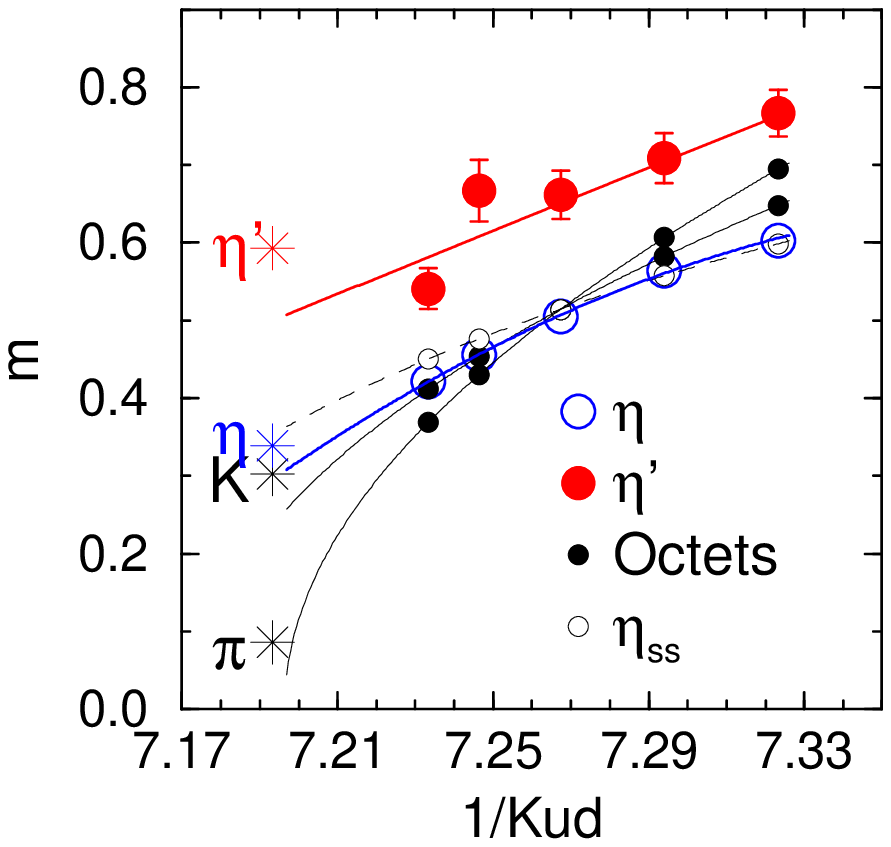}
  \caption{PS meson masses including $\eta^\prime$ and $\eta$ as a
 function of $1/K_{\rm ud}$ at $K_{\rm s}=0.13710$(Left) and 0.13760(Right).}
  \label{fig:singlet}
\end{figure}
In this subsection, I briefly present a preliminary result on the flavor singlet meson mass\cite{yoshie1}.
In Fig.~\ref{fig:singlet} we present PS meson masses including
$\eta^\prime$ and $\eta$ as a function of the light quark mass 
($1/K_{\rm ud}$ where $K$ is the hopping parameter) for 2 values of the
strange quark mass ($1/K_{\rm s}$) at $a\simeq 0.12$ fm. Small solid
circles denote LL and LS meson results while small open circles
correspond to SS meson results without disconnected diagrams. Once we
correctly include contributions from disconnected diagrams, the SS state is
mixed with the flavor singlet state, so that the mass of the SS state
becomes a little lighter, as shown by large open 
circles in the figure.
The singlet $\eta^\prime$, denoted by large solid circles in 
the figure, appears much heavier than other PS mesons.

By the polynomial chiral extrapolation to the physical point, we obtain 
$m_\eta = 545(16)$ MeV, consistent with the experimental value (550 MeV), while
$m_{\eta^\prime} = 871(46)$ MeV, which is much larger than octet PS
meson masses and is smaller than the experimental value (960 MeV) only
by 100 MeV (2 $\sigma$). The U(1) problem seems to be solved. More
studies at two other lattice spacings, however, will be required for the 
final conclusion.

\section{Summary and outlook}

\subsection{Summary}
CP-PACS and JLQCD collaborations has performed the 2+1 full QCD project, using the RG improved gauge action and non-perturbatively $O(a)$ improved clover quark action. Configuration generations have already been completed and the analyses are now being finalized. Light meson masses agree with experimental values after the continuum extrapolation assuming that the $a^2$ contribution dominates the scaling violation.  Values of the up-down quark mass and the strange quark mass are determined in the continuum limit. We observe that the dynamical strange quark effect is much small than that of the up-down quarks.

Currently there are several on-going analyses, which include  the
non-perturbative determination of renormalization factors 
to remove an ambiguity of 1-loop estimates,
the flavor singlet meson mass as  
presented, and heavy quark quantities using a relativistic heavy quark action.

\subsection{PACS-CS project}
We have just started a new project, PACS-CS project, which uses a new cluster PACS-CS.
The PACS-CS starts operating this July with the peak speed of 14.3 Tflops\cite{ukawa1}. 
In order to remove the most serious ambiguity due to the chiral 
extrapolation,
the PACS-CS collaboration wishes to go down to lighter up-down quark 
masses with the clover fermion, employing the domain decomposed HMC 
algorithm proposed by L\"usher\cite{luescher1}. Our preliminary test 
study indicates that we can go down to as small as 15 MeV quark 
mass\cite{kuramashi1}.

\begin{figure}
  \includegraphics[width=5cm, angle=270]{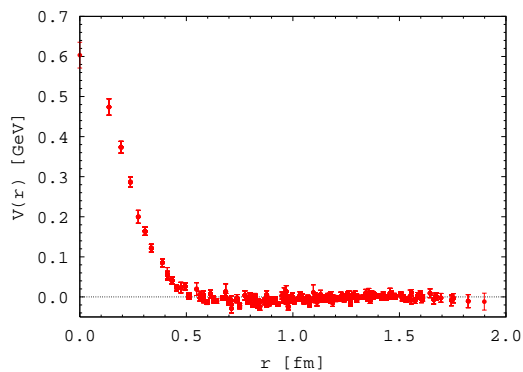}
  \caption{The $N$ $N$ potential.}
  \label{fig:Nforce}
\end{figure}

\subsection{Nucleon force}
Last but not least, 
I briefly introduce an interesting application of lattice QCD technique to
the nucleon force (the potential between two nucleons). Recently we try to extract this $N$ $N$ potential on the lattice from the Bethe-Salpeter wave function $\phi$ and the effective Schr\"odinger equation as
\[
V(r) = E +\frac{1}{m_N} \frac{\nabla^2 \phi (r)}{\phi (r)} .
\]
Fig.~\ref{fig:Nforce} gives the $N$ $N$ potential for the 
$(J^P,I)=(0^+,1)$ channel,  obtained in quenched QCD at 
$a\simeq 0.14$ fm and $m_\pi\simeq 880$ MeV\cite{ishii1}.  We clearly 
observe the strong repulsive force at the short distances. 
Although errors are still too large to see an expected attractive force 
at the intermediate distance, this method seems promising. 
Currently we investigate systematics of this method.




\begin{theacknowledgments}
I am grateful to all members of CP-PACS and JLQCD collaborations. 
In particular  I would like to thank Drs. N. Ishii, T. Ishikawa,
 Y. Kuramashi and T. Yoshi\'e  for providing me data and figures used 
in this talk. This work is supported in part by  Grant-in-Aid of the 
Ministry of Education (Nos. 13135204,15540251). 
\end{theacknowledgments}



\bibliographystyle{aipproc}   

\begin{thebibliography}{99}

\bibitem{cppacs1}
CP-PACS Collaboration: S.~Aoki {\it et~al.},
\emph{Phys. Rev. Lett.} \textbf{84}, 238-241 (2000);
\emph{Phys. Rev. } \textbf{D67}, 034503 (2002).

\bibitem{cppacs2}
CP-PACS Collaboration: A.~Ali Khan {\it et~al.},
\emph{Phys. Rev. Lett.} \textbf{85}, 4674 (2000); \emph{Eratum-ibid.} \textbf{90}, 029902 (2003);
\emph{Phys. Rev.} \textbf{D65}, 054505 (2002); \emph{Eratum-ibid.} \textbf{D67}, 059901 (2003).

\bibitem{ishikawa1}  CP-PACS and JLQCD Collaborations: T.~Ishikawa {\it et~al.},
\emph{Nucl. Phys. } \textbf{B}(Proc. Suppl.)\textbf{140}, 225 (2005).

\bibitem{ishikawa2} CP-PACS and JLQCD Collaborations: T.~Ishikawa {\it et~al.},
\emph{PoS } \textbf{LAT2005}, 057 (2005).

\bibitem{ishikawa3} CP-PACS and JLQCD Collaborations: T.~Ishikawa {\it et~al.},
\emph{PoS } \textbf{LAT2006}, 181 (2006).

\bibitem{joint1} CP-PACS and JLQCD Collaborations: S.~Aoki {\it et~al.},
\emph{Phys. Rev. } \textbf{D73}, 034501 (2006).

\bibitem{joint2} JLQCD Collaboration: S.~Aoki {\it et~al.},
\emph{Phys. Rev. } \textbf{D65}, 094507 (2002).

\bibitem{Aoki:1998-PT-renorm}
S.~Aoki {\it et~al.},
\emph{Phys. Rev.} \textbf{ D58}, 074505 (1998).

\bibitem{aoki1}
S.~Aoki,
\emph{Phys. Rev.}\textbf{D68}, 054508 (2003).

\bibitem{chpt1}
S.~Aoki, O.~B\"ar, T.¡ÁIshikawa and S.~Takeda,
\emph{Phys. Rev.} \textbf{D73}, 014511 (2006) .

\bibitem{chpt2}
S.~Aoki, O.~B\"ar and S.~Takeda,
\emph{Phys. Rev.} \textbf{D73}, 094501 (2006) .

\bibitem{yoshie1}
S.~Aoki {\it et~al.},
\emph{PoS} \textbf{LAT2006} (2006) (hep-lat/0610021).

\bibitem{ukawa1}
PACS-CS Collaboration: A.~Ukawa {\it et~al.},  
\emph{PoS} \textbf{LAT2006},  039 (2006).

\bibitem{luescher1}
\emph{JHEP} \textbf{ 05}, 052 (2003);
\emph{Comput. Phys. Commun.} \textbf{156}, 209 (2004);
\emph{ibid} \textbf{165}, 199 (2005).

\bibitem{kuramashi1}
PACS-CS Collaboration:  Y.~Kuramashi {\it et~al.}
\emph{PoS} \textbf{LAT2006},  029 (2006).

\bibitem{ishii1}
N.~Ishii, S.~Aoki and T.~Hatsuda,
\emph{PoS} \textbf{LAT2006},  (2006) (hep-lat/061002).


\end{thebibliography}

\end{document}